%
\magnification=\magstephalf
\def\pp{\noindent\parshape 2 0truecm 15truecm 2truecm 13truecm}
%
\def\apjref#1;#2;#3;#4 {\par\pp#1, {\it #2}, {\bf #3}, #4. \par}
%
%

%
%
\def\np{\vfill\eject}
%
%
\vsize=22.5 true cm
\hsize=16.0  true cm
\parskip=0.2cm
\parindent=20pt
\raggedbottom
\pretolerance=10000
\tolerance=10000
\def\etal{\sl et al.~\rm}
%
\def\ref#1{\global\advance\fcount by 1 \global\xdef#1{\relax\the\fcount}}
\def\pp{\parshape 2 0truecm 15truecm 2truecm 13truecm}

\def\simlt{\lower.5ex\hbox{$\; \buildrel < \over \sim \;$}}
\def\simgt{\lower.5ex\hbox{$\; \buildrel > \over \sim \;$}}
\def\boxit#1{\vbox{\hrule\hbox{\vrule\kern3pt
      \vbox{\kern3pt#1\kern3pt}\kern3pt\vrule}\hrule}}

\def\myfig#1#2#3#4{
  \midinsert
     \vskip #2 true cm \vskip 1.3 true cm
     #3
     \vskip -2. true cm
     {\baselineskip 11pt
       \rightskip=0.8 true cm \leftskip=0.8 true cm
       \par\noindent{\bf Figure #1:}  #4 \par
       \rightskip=0 true cm \leftskip=0 true cm
     }
     \vskip -.1 true cm
  \endinsert
}

\def\ifm#1{\relax\ifmmode#1\else$\mathsurround=0pt #1$\fi}
\def\Msun{~{\rm M}\ifm{_\odot}} 
\font\bigbf=cmbx12 at 14.4truept
 at 10.0truept
\input psfig
\overfullrule=0pt

\centerline {\bigbf A Galaxy-Weighted Measure of the Relative}
\centerline {\bigbf Peculiar Velocity Dispersion}
\bigskip\bigskip

\centerline{Marc Davis}
\centerline{Depts. of Astronomy and Physics, Univ. of
California, Berkeley}
\centerline{marc@coma.berkeley.edu}

\medskip
\centerline {Amber Miller} 
\centerline{Dept. of Physics, Princeton University}
\centerline{ amber@pupgg.princeton.edu}

\medskip
\centerline{ and}
\medskip
\centerline{ Simon D.M. White}
\centerline { MPI fuer Astrophysik, Garching}
\centerline{  swhite@mpa-garching.mpg.de}

\bigskip
\centerline {\bigbf Abstract}
\medskip
 The relative pair dispersion of galaxies has for the past decade
been the standard measure of the thermal
energy  of fluctuations in the observed galaxy
distribution.  This statistic is known to be unstable, since it
is a pair-weighted measure that is very sensitive to rare,
rich clusters of galaxies.  As a more stable alternative, we here
present a  single-particle-weighted statistic $\sigma_1$, 
which can be considered as an estimate of the one-dimensional rms peculiar
velocity dispersion of galaxies relative to their neighbors, and 
which can be interpreted by means of 
a filtered version of the Cosmic-Energy equation.  We
calculate this statistic for the all-sky survey of IRAS galaxies,
finding $\sigma_1=95 \pm 16 $ km/sec. The UGC catalog yields a higher 
value, 
$\sigma_1=130 \pm 15$ km/s.  We calibrate our
procedure by means of mock catalogs constructed from N-body
simulations and find that our method is stable and has modest
biases which can easily be corrected.
We use the measured values of $\sigma_1$
 in a filtered Layzer-Irvine equation to obtain an
estimate of $\tilde\Omega \equiv \Omega/b^2$. 
We find that $\tilde\Omega\approx 0.14 \pm 0.05$ for both
the IRAS and UGC catalogs,  which is slightly lower than other recent
determinations, but is consistent with a trend of
an effective $\Omega$ that increases gradually with scale. 

\np
\centerline {\bigbf 1. Introduction}
\medskip

Considerable effort has been devoted in the past decade to the
measurement of the quantity $\sigma_{12}(r)$, the relative
peculiar velocity dispersion of pairs of galaxies as a function
of their separation (Davis et al 1978; Peebles 1979, 1981;
Davis and Peebles 1983, hereafter DP83; Bean et al  1983;
de Lapparent et al 1988; Hale-Sutton et al 1989; Mo et al 1993;
Zurek et al 1994;  Fisher et al 1994a; Marzke et al 1995; Brainerd et al 1996,
Somerville \etal 1996).  
The relative pair velocity dispersion is most easily
extracted from the two-point correlation function in redshift
space, $\xi(r_p,\pi)$. Since this correlation
is a pair-weighted quantity, so is the rms
peculiar velocity dispersion. Dense clusters of galaxies
contain many pairs and have  high internal velocity dispersion. 
Consequently they can dominate the measured
dispersion, as shown by Mo et al. (1993), Zurek et al. (1994), and
Somerville \etal (1996), 
who found that the elimination of the cores of dense clusters leads to
a significantly reduction in the measured $\sigma_{12}$ and argued
that the ``true'' value might be substantially larger than earlier
estimates. For these reasons, many authors (Hale-Sutton et al
1989; Mo et al.  1993; Zurek et al. 1994; Guzzo et al. 1996)
have concluded that this statistic is unreliable. It is possible that
a reliable estimate has finally been achieved with the nearly complete
CfA2 survey (Marzke et al 1995), but until these
results are confirmed by even larger surveys,  a cloud of uncertainty
will remain.

The relative pair dispersion $\sigma_{12}(r)$ is well defined
from the point of view of kinetic theory (Davis and Peebles 1977,
Peebles 1980) and is an essential ingredient in the cosmic virial
theorem (CVT), which is itself a statement of equilibrium between
kinetic pressure and gravitational acceleration averaged across
all virialized systems. The proper evaluation of the cosmic virial
theorem includes a nearly divergent integral over the poorly
constrained 3-point correlation function of galaxies $\zeta$ (see
Peebles 1980, eq. 75.10).  Comparison of  $\sigma_{12}$ to this
integral yields a measure of the cosmic density parameter $\Omega$,
subject
to all the usual arguments about bias in the galaxy distribution.  The
sensitivity of  $\sigma_{12}$ to the treatment of the
rare dense clusters is very likely matched by similar sensitivity
in the integral over $\zeta$, but this has never been checked
directly.  Juszkiewicz and Yahil (1989) have shown  how the standard
CVT, which
applies to nonlinear clustering on small scales, can be readily 
extended to the linear regime valid on large
scales. Mo \etal (1996) have shown how analytical approximations
to $\sigma_{12}$ and to other low order statistics can be derived
from the initial linear power spectrum $P(k)$.  
 Bartlett and Blanchard (1993, 1996) have pointed out that the 
integral of the three-point function $\zeta$ should be taken over
the {\it galaxy-galaxy-mass} correlations, and that the usual estimate
of $\zeta$ based on 
{\it galaxy-galaxy-galaxy} correlations can seriously underestimate
the inferred $\Omega$, if galaxies have extended massive halos.
Kepner \etal (1996) have suggested that much of sampling variance of
$\sigma_{12}$ is likely to be reduced if it is tabulated as a function of
the galaxy density surrounding each pair.  

Although the CVT test for $\Omega$ is rarely performed with
any rigor, $\sigma_{12}(r)$ for the mass distribution is easily
measured in high resolution N-body simulations, and is often used to
compare them with observation (e.g. Davis et al 1985).  The fact that the
observed pair dispersion in the galaxy distribution is well below that
expected for the mass distribution in an $\Omega=1$ cosmology was the
major factor motivating the concept of bias in the galaxy distribution.
Considerable attention has been given to the distinction between
$\sigma_{12}(r)$ measured for galaxies and the same quantity measured 
for the mass in simulations (e.g.  Davis et al 1985; Couchman et al
1990; Gelb and Bertschinger 1994; Zurek et al 1994).
The results are, not surprisingly, very sensitive to the spatial
resolution of the simulation and to the degree to which clumps of
dark matter in the simulations are separated into distinct ``galaxies''.

There is clearly a need for a simple statistic to evaluate the kinetic
energy of the galaxy distribution, preferably one that is both easily
evaluated and stable.    In this paper, we develop a statistic
which contains much of the same information as $\sigma_{12}$ but
is  galaxy-weighted rather than a pair-weighted.  We describe the
statistic and its theoretical underpinning in Section 2, and
apply it to several redshift survey catalogs in Section
3.  We demonstrate the robustness of our statistic by applying
it to a series of mock catalogs extracted from a high resolution
N-body simulation.  
We believe that it will prove both robust and useful in
diagnosing the thermal state of the galaxy distribution.

\bigskip
\centerline {\bigbf 2. Cosmic Energy}
\medskip
\noindent{\bf 2.1. The Standard Layzer-Irvine Equation}

Consider the cosmic energy equation, also known as the Layzer-Irvine
equation, which describes the relationship between the kinetic
and potential energies of the fully-nonlinear fluctuation field.
Consider a sample with mass $M = \sum m_i = \rho_b V$ where $\rho_b = 
3\Omega H_o^2/8\pi  G$ is the mean matter density.  
The specific kinetic $K$ energy in fluctuations can be written as a sum 
over the particles 
$$ K = {\rho_b\over 2M} \int d^3x(1+\delta({\bf x}))(\bar{\bf v}^2({\bf x}) +
\sigma^2({\bf x})) ~~ = ~~
{3 \over 2}\langle  v_p^2\rangle \eqno(1)$$ 
where $\delta({\bf x})$, $\bar{\bf v}({\bf x})$ and $\sigma({\bf x})$ 
are the overdensity and the mean and dispersion in
peculiar velocity of the particles at position ${\bf x}$, and 
$<v_p^2>^{1/2}$ is the one-dimensional peculiar velocity
dispersion averaged over all particles.
The specific potential energy $W$ is given by 
$$ W =  -{Ga^2 \rho_b^2\over 2M} 
\int d^3x_1d^3x_2 \delta({\bf x_1})\delta({\bf x_2})/x_{12} ~~ = ~~
-2\pi G\rho_b J_2 \eqno(2)$$ 
where $x_{12}=|{\bf x_1} - {\bf x_2}|$ and $J_2$ is given by the usual
expression, 
$J_2=\int\xi(r)rdr$ with $\xi(r)$ the two-point correlation
function of the mass distribution. Thus
$${W =-{3\over 4}\Omega H_0^2J_2.} \eqno(3)$$
The Layzer-Irvine  equation is an exact result relating
 the time evolution of $K$ and
$W$, 
$$ {{dK\over dt}+{dW\over dt}+{{\dot a}\over a}(2K+W)=0}\eqno(4)$$
(Peebles 1980 eq. [24.7]), where $a$ is the expansion parameter.

In the limit of self-similar clustering, equation (4) reduces to an algebraic 
expression  (Peebles 1980 eq.
[74.6], DP83 eq. [33]), 
$$ K = {4\over 7+n}|W|  \eqno(5)$$
where $n$ is the index of the linear power spectrum, $P(k)\propto
k^n$. In applications to real data one might take $n \approx -1$,
the effective slope of the power spectrum on the scale of nonlinear 
clustering, i.e. near $8$ h$^{-1}$ Mpc.  Note that for $n=-1$ equation
(5) gives $K/|W| = 2/3$, which, as we show below, is the value
expected for linear clustering in an Einstein-de Sitter universe with 
any power spectrum. In an open universe with no growing
modes and only stable virialized clusters, the Layzer-Irvine
equation becomes $K/|W|\approx 1/2$.  Thus we can encompass the likely range
of possibilities for our Universe by writing Eq (5) as
$$ \langle v_p^2\rangle \approx g\Omega H_0^2J_2  \eqno(6)$$
where $g$ is in the restricted range $1/4 < g < 1/3$.

This form of the Layzer-Irvine equation was first used by Fall (1976)
to set constraints on $\Omega$ based on the Shapley-Ames redshift
catalog.  If we approximate $\xi(r)$ by a power law,
$\xi(r) = ({r\over r_0})^{-\gamma}$ 
with a cutoff at $r> x r_0$,
then we expect a one-dimensional rms peculiar velocity
$$ \langle v_p^2\rangle^{1/2}  = \Omega^{1/2} \left(g
x^{2-\gamma} \over (2 - \gamma)\right)^{1/2} H_0 r_0 \approx 1.4
\Omega^{1/2} H_0 r_0 \eqno(7)$$
where the last equality results from $\gamma=1.8$, $x=4$, and 
$g=1/3$.  Thus in an $\Omega =1$ Universe in which galaxies
trace the mass with $H_0 r_0 = 500$ km/s, the 1-d rms velocity of
particles relative to the comoving frame
is expected to be quite large, 700 km/s,  corresponding to an {\it rms} 
three-dimensional velocity of 1200 km/s.  This can be compared with the 
peculiar velocity of our own galaxy, 620 km/s.

There are several  well known problems with equation (7) that
have prevented its widespread application in cosmology. First note that the
integrand of $J_2$ is divergent at large scale for a power law
correlation function; the convergence of the integral is dependent on the 
uncertain turnover scale from pure power law behavior, 
which is  only weakly constrained, and 
therefore $J_2$ is poorly determined.  Furthermore, 
$\langle v_p^2\rangle^{1/2}$ is the rms velocity of matter relative to the
comoving frame, but this quantity 
can be reliably measured only 
for the Milky Way.  The Layzer-Irvine equation is a very simple
statistic that applies on a global scale; it is an average of the
energy balance across all scales. The kinetic energy term
includes nonlinear motion within groups as well as large scale,
coherent flows.  Similarly, 
the potential term is an integral of the fluctuations
on all scales both large and small. The problem in the
application of the cosmic energy equation is that 
reliable measurements of the largest scale contributions are not
available for either the
kinetic or potential energy terms.

\bigskip
\noindent{\bf 2.2. A Filtered Version of the Cosmic Energy Equation}

The problems with the cosmic energy equation can be overcome if we
consider a filtered version of both K and W. 
The Fourier decompositions of the density contrast and the mean peculiar 
velocity are
$$ \delta({\bf x}) = {1\over (2\pi)^3} \int d^3k~ \delta_{\bf k} 
e^{i{\bf k\cdot x}}~~,$$ 
$$ {\bf \bar{v}}({\bf x}) = {1\over (2\pi)^3} \int d^3k~ {\bf \bar{v}_k} 
e^{i{\bf k\cdot x}} ~~.	\eqno(8) $$
The large-scale but small amplitude fluctuations in density which make
it difficult to evaluate $J_2$ are represented by the small $k$
Fourier components $\delta_{\bf k}$ and the large-scale streaming motions
to which they give rise are represented by the small $k$ components
${\bf \bar{v}_k}$. According to 
late-time linear theory (Peebles 1980) these two quantities are related by
$$ {\bf \bar{v}_k} = H_0 f(\Omega) {\delta_{\bf k} \over i{\bf k}}~,\eqno(9)$$
where to a good approximation $f(\Omega)=\Omega^{0.6}$. The
corresponding contributions to $W$ and $K$ can be written as
$$ W_k = {Ga^2 \rho_b \over (2\pi)^2 V} {|\delta_k^2|\over k^2}
\eqno(10)$$
and
$$ K_k = {1\over 2 V (2\pi)^3}  |{\bf v_k}|^2 ~, \eqno(11) $$
so that in linear theory $K_k$ and $W_k$ are related by
$$ K_k = {2 f^2(\Omega)\over 3 \Omega} |W_k|~. \eqno(12) $$
Since this is true for each linear mode, we see that
$$K/|W|\approx 0.667 \Omega^{0.2}\eqno(13) $$
holds for the entire large-scale linear contribution to  
the Layzer-Irvine equation.

This suggests that we consider a filtered version of equation (6)
with these uncertain large-scale contributions removed:
$$ {\tilde v_p^2\approx {\tilde g}\Omega H_0^2 {\tilde J_2} } \eqno(14)$$
where $\tilde v_p^2$ and ${\tilde J_2}$ are high spatial frequency 
versions of the rms peculiar velocity and potential energy and ${\tilde g}$
is a modified scaling constant that could be somewhat scale-dependent.  

To be safe we would like to remove large-scale contributions only on
scales where we are sure the distribution is linear.
However, with available data sets, it is necessary to filter on 
relatively small scales to have acceptable signal to noise in the
resulting statistics.  To test how well equation (14) works as a
function of filtering scale, we examined the output from several high
resolution PPPM N-body simulations.  We have studied one simulation with
$n=-1$ power law initial fluctuations and $\Omega=1$, and another with
$n=-1$ and $\Omega=0.1$, both with $10^6$ particles and with
size of $\approx 200h^{-1}$ Mpc.  We have also
examined a simulation of $2\times 10^6$ particles
 with CDM initial conditions and
$\Omega=1$.  Table 1 gives values $K/|W|$ as a function of filtering
scale for these simulations.  We use a gaussian filter to smooth
the potential energy in Fourier space and the velocities in real space.
We give the scale of the gaussian
filter, $\sigma_s$, in units of the matter correlation length, $r_0$, and we
list the kinetic energy density in small scale motions for the
$\Omega=1$, 
$n=-1$ model (in arbitrary units). Finally for each model we give the ratio
$K/|W|$ both for the high frequency structure retained in equation
(14) and for the large-scale contributions which have been removed.
Notice that for both $n=-1$ models the large-scale ratio approaches
the value expected from equation (13) as the smoothing is increased,
but that the CDM model is still far from the linear prediction on the
largest scale considered. 

One expects ${\tilde g}$ to be scale-dependent because on sufficiently
small scale the filtered kinetic energy will be larger than the filtered
potential energy, implying merely that the systems are not bound on that scale.
This is equivalent to the situation within galaxies:  stars orbiting within a
10 kpc elliptical galaxy are not bound by the mass distribution on a 1 kpc
scale.  On the other hand, ${\tilde g} \approx 1/4$ is a likely lower bound
consistent with virial equilibrium on small scales, and thus can provide
an upper limit to the derived density $\Omega$.
The physical interpretation of the filtered cosmic energy equation is
simply that there is a balance between
potential and kinetic energy provided all scales are included up to
those of the largest virialized systems.   
The great advantage of the filtering is that, by deleting the
long wavelength modes, the kinetic energy term
can be  measured by the motion of galaxies relative to their neighbors;
common large-scale motions need not be considered.  The filtered value
of W is easily calculated as a suitably filtered version of the
correlation function integral $J_2$.  The small-scale values of
$K/|W|$ in Table 1 are only slightly model-dependent, so there is
little ambiguity in applying equation (14).

In the next section,
we show how to construct a suitably filtered velocity dispersion.  
We proceed by constructing
the relative distribution of pairs because this can be done using
redshift information alone and automatically filters out common
motions.  To prevent this statistic from being pair weighted as in the 
$\sigma_{12}$ analysis, 
we shall construct a mean distribution function of relative velocities
in which each galaxy is given equal weight.

\bigskip
\noindent {\bigbf 3. A Galaxy Weighted Velocity Statistic}
\bigskip 
\noindent {\bf 3.1 The Samples}
\medskip

The primary data we shall use comes from an all sky redshift survey of
galaxies from the Infrared Astronomical Satellite (IRAS) data
base, flux limited to 1.2 Jy at 60$\mu$ (Fisher et al 1995).  We
semi-volume limit the sample by deleting galaxies with 60$\mu$
flux too low for them to be included in the sample if they are
placed at a redshift of $4000$ km/sec, and we truncate the sample
at redshift 8000 km/s. The IRAS sample thus selected contains
2374 galaxies.  We do not consider cosmic flow fields in
constructing this subsample, presuming redshifts in the LG frame to be
equivalent to distances.  We also examine a sample of 1959
optical galaxies taken from the Uppsala General Catalog (UGC)
with $m < 14.5$ and $b > 30^\circ$, using the same volume limit
and redshift limit as for the IRAS catalog.  This UGC catalog is
a subsample of the recently completed ORS redshift survey
(Santiago et al 1995, 1996).

\bigskip
\noindent {\bf 3.2 Building the distribution of relative velocities}
\medskip

Consider a galaxy in the catalog.  We consider any other galaxy
with projected separation less than a limiting value $r_p$ and with
a redshift separation of less than 1200 km/s to be a neighbor of the
original galaxy.  
The distribution $P$ of observed
$\delta z$ as a function of pair separation is constructed
for each galaxy individually along with the smoothed expected background 
distribution based on the selection function and density for
each sample. 
Thus if the catalog's selection function is $\phi(v)$ and the mean density is
$\bar n$, then the expected smooth background at redshift separation 
$\pm\Delta v$ 
from a galaxy at measured redshift $v$ is simply 
$$B(\Delta v) = f_B \pi r_p^2~ \delta v \bar{n} \phi(v \pm \Delta v) 
 \eqno(15)$$
where $f_B$ is a fraction of the area of the projected circle
that falls within the catalog boundaries, and $\delta v = 50$
km/s is
the width of the binning in redshift space. 
This background distribution $B$ is then subtracted from the
pair distribution $P$ and the resultant distribution is normalized by the 
difference between the total number of real pairs and the total number of pairs
expected from a the background distribution.  
That is, for each galaxy we construct a statistic $G$ given by
$$ {G(\Delta v)={P(\Delta v)-B(\Delta v) \over 
\sum_{\Delta v} (P(\Delta v) -  B(\Delta v))}} \eqno(16)$$
where the summation is taken over the bins of redshift separation.

The total distribution, $D(\Delta v)$, is built by adding together the
individual galaxy distributions, $D(\Delta v)=\sum_n G_n(\Delta v)$. 
This procedure
normalizes the distribution of pairs around each galaxy by the
total number of excess pairs for that galaxy, rather than equally weighting
all galaxy pairs.  Only galaxies for which the total number of pairs exceeds
the total background by at least one are included in the summation defining
$D$.  By deleting the galaxies with fewer pairs than expected at random,
we bias the distribution toward
the denser regions, but this bias can be readily corrected.  An alternative
might be to accumulate the normalized $P(\Delta v)$
statistic, with no individual background subtraction,  
but then the background subtraction from $D(\Delta v)$ would be
problematic since the expected background
is not the same for each galaxy in the catalog.

The resulting distributions for the IRAS, UGC,  and
(uncooled) N-body data sets are plotted in Figure 1. These curves
simply represent the probability that a random galaxy has excess
neighbors with
a given projected separation and redshift separation (velocity difference)
 $\Delta v$.  Since
we know that galaxies are correlated in real space, we would expect this
distribution function to have a non-zero width even in the absence
of redshift space distortions.  Interpretation of this distribution 
function in terms of random peculiar  motions is described below.

\bigskip 
\noindent {\bf 3.3 Mock Catalog Samples}
\medskip

In order to gain a better understanding of what is being measured
in this procedure,  we 
make extensive use of an N-body mock catalog generated from a
high resolution N-body simulation of $10^6$ particles with $\Omega =
1$ and with power law spectral index $n=-1$.  The method by which the
mock  catalogs are generated is discussed in Fisher et al (1994b)
and Davis, Nusser, Willick (1996).  This sample, which is 
unbiased relative to the mass, is similarly
flux limited and contains 4092 ``galaxies".
 We have chosen a scaling of the
simulation such that the correlation 
length $r_0$  closely matches
that of the IRAS catalog, $r_0 \approx 3.6h^{-1}$ Mpc.

We shall find that the measured value of the dispersion obtained
for the N-body catalog differs significantly from that obtained
for the IRAS and UGC catalogs.  This is not unexpected because it
is known that the N-body simulations produce small-scale velocity
fields much hotter than those observed. In order to produce
models that mimic the observations, the velocity fields in these simulations
are often smoothed using a Gaussian window function (Fisher et
al 1994b).  In the present case, however, it is important not to alter the
distribution function of peculiar velocities on small scales. 
Instead of smoothing the velocity field, we transform the 
N-body galaxies to real space using the known individual 
peculiar velocities, and
then divide those peculiar velocities in half 
before transforming back into redshift space. This procedure
has the advantage that it reduces the velocity dispersion
without changing the shape of small-scale distribution.  
Thus we have both a hot
and a cooled version of the same mock catalog which we shall use in
the analysis below. 
Although this is admittedly not a self-consistent procedure for
constructing the Universe, it does provide a fair test of our ability to
recover the amplitude of the small-scale velocity field in the presence
of strong clustering.

 We also examined a series of mock catalogs which were cooled in the
fashion described by Davis, Nusser, and Willick (1996), in which
a smoothed version of the velocity field is averaged with the original
velocity field.  This has the effect of preserving the original amplitude
of the large scale flows and of diminishing the small scale velocities
by a factor of two.  We find that this more sophisticated cooling
yields very similar results to the simpler cooling procedure.

The N-body models can also be used to demonstrate the separation of
large and small-scale peculiar motions.  Our velocity statistic is
sensitive to the small-scale peculiar motions, the motion of neigboring
galaxies relative to each other.  But a substantial fraction of the
peculiar motion of galaxies is coherent on small scales and is
a common-mode, bulk flow motion that is filtered out by our procedure.

To study this separation,  we make use of the true
peculiar velocity information provided by the mock IRAS, N-body catalogs to
transform the mock samples from redshift space to real space. 
  Consider the set of ``galaxies" that
have excess neigbors within a  sphere in real
space of  radius $r_s = 2 h^{-1}$  Mpc.  
The solid curve of Figure 2 shows the peculiar velocity distribution 
of these points, relative to the comoving frame of the simulation.  Note
that the distribution is asymmetric, with  long tails.  The asymmetry 
is a consequence of the anisotropic distribution of the galaxies in the mock
catalog, which has been carefully designed to mimic the actual IRAS survey with
its large-scale flows and large motion (600 km/s) of the central observer.
The dashed curve in Figure 2 shows the peculiar 
velocity distribution of these same points
relative to the mean peculiar velocity of  their  neighbors within the same
sphere.  The common-mode, bulk flow motion has been removed, 
and the distribution is now symmetric about zero, with a much narrower
width and modest tails.

For a larger neighbor window, we might expect the internal
velocity dispersion to increase, while the rms of the external
(or bulk flow)
peculiar velocity dispersion should decrease.
We examine ``galaxies" with neighbors within  real space spheres
of radius ranging from
$1h^{-1}$ Mpc to $5h^{-1}$ Mpc with results for the internal and bulk flow
dispersion given in Table 2. These dispersions do NOT behave in the
naively expected manner;
we find that the rms of the external (bulk flow) peculiar velocity
dispersion does decrease with increasing scale but so does the
internal velocity dispersion, 
in contrast to the behavior of the simulation as a whole,
as given in Table 1.  

The fourth column in Table 2 is the true rms value of the
peculiar velocity (relative to the 
comoving frame) for all the points having neighbors within the
selection radius.  This column provides an explanation of the
behavior of the dispersions measured 
as a function of scale. We see
that the galaxies entering into our analysis are not a fair
sample of the total population as measured by the rms
dispersion.   The galaxies with pairs on
increasingly small scales represent a more and more biased
subsample of all the points; the selected `galaxies' with
pairs are preferentially found in the denser regions, which have 
higher peculiar velocities. This effect can also
be seen by making maps of the galaxies on ``redshift shells'',
maps of the galaxies on the sky which have redshifts between two
limiting values.  Galaxies with peculiar velocities larger than
$1000$ km/sec tend to be located within the denser regions of the
maps.  
Because we are working
with flux limited catalogs, galaxies at higher redshifts having
pairs are even more biased than galaxies at lower redshift.

The bias so induced can be estimated from the fifth and sixth columns
of Table 2.  Here we list the 
fraction of mock points having real space neighbors within a given scale, 
  along with the ratio of
the rms peculiar velocity of these points relative to
all the points in the mock catalog.  Note that less than half of
the ``galaxies" have neighbors within $2h^{-1}$ Mpc and that
these points have slightly higher than average velocity dispersion.

Since the true peculiar velocities are not known for
the IRAS or UGC galaxies, these catalogs cannot be converted to real space.
Therefore for direct comparison with the IRAS or UGC distributions, 
it is essential to go back to
redshift space. In redshift space, galaxies
having fewer neighbors than expected from a random background are
excluded from analysis as described above.
Table 3 shows the fraction of galaxies included in the
analysis for each catalog. The Table also shows
the ratio of the rms dispersion of galaxies in the analysis to the rms
dispersion of all the galaxies in the catalog. This latter
quantity can be directly measured only for the mock catalog.

Fortunately a nearly identical fraction of galaxies within the IRAS, UGC, 
and the N-body samples have neighbors in excess of the random
background, and  we shall therefore 
assume that the ratio of the dispersions behave similarly.
We thus shall decrease the measured
value of the IRAS and UGC dispersions by a
factor of $1.19$, the value obtained for the mock catalog in Table 3.

\bigskip
\noindent {\bf 3.4 Measurement of the intrinsic dispersion, $\sigma_I$}
\medskip 

For the analysis in redshift space, we have chosen pairs to be
galaxies separated by up to $2$ h$^{-1}$ Mpc projected separation and $1000$
km/sec along the line of sight.  The distribution $D(\Delta v)$ describes
the distribution function the radial velocity of neighbors in excess of random,
summed over all galaxies in the sample.
Because galaxies are clustered 
the second moment of the distribution of $D$ would be
nonzero even in the absence of peculiar velocities.  
The density of neighboring galaxies expected in excess of a randomly
chosen galaxy is simply proportional to the  
two point correlation function $\xi(r)$.  
In order to interpret the measured $D(\Delta v)$,  we note that
the observed redshift space correlation function is a convolution
of the true spatial correlation function $\xi(r)$ with a
velocity distribution function, which we can write as  
$$ \xi (r_p,\pi) = C\int {dy \over \sigma_I(r)}\xi(r)
{\rm exp}\biggl(-\eta \bigg | {\pi -y \over \sigma_I (r) }
\bigg |^\nu\biggr) \eqno(18)$$
where $r=\sqrt{r_p^2+y^2}$ (Fisher \etal II eq. [12]). Fisher \etal
note that both the pair weighted velocity distribution functions for the
IRAS data and the N-body models are adequately
fit by an exponential model  ($\eta =
\sqrt{2}$, $C=1/\sqrt{2}$, and $\nu =1$). There is no guarantee,
however, that an exponential model will yield the best fit to the
new particle weighted distribution function, and we
have tested gaussian models ($C=1/\sqrt{2\pi}$, $\eta=1/2$, and $\nu=2$)
as well.  With these choices of $\eta$, $\sigma_I$ is
the rms dispersion of galaxies relative
to their neighbors in both the exponential and gaussian models.
We define  $\sigma_I(r)$ to be the intrinsic dispersion, and set it to 
be constant in
$r$.  We furthermore ignore streaming effects of the sort discussed
by  DP83 and Fisher \etal II, since  
 we are interested in the mean velocity dispersion around individual galaxies, 
not the mean 
streaming and velocity dispersion of pairs of galaxies.
Thus we construct a model velocity distribution function $M(\pi)$  as 
$$ { M(\pi) = {2 \pi C_1 \over\sigma_I}
 \int_0^{2h^{-1}} dr_p r_p \int_{-\infty}^\infty dy ~\xi(r) 
 {\rm exp}\bigg(-{\eta \over \sigma_I}\big| {\pi -y}\big|^{\nu} \bigg).}
\eqno(19)$$

The normalizing constant $C_1$  depends on the selection function and
redshift distribution of
the galaxies with neighbors in the sample but it will scale out of the 
analysis.  Thus only the shape, and not the amplitude, of two point correlation
function is required for the modeling of $D(\Delta v)$. 
For the IRAS and UGC data, the function, $\xi (r)$ is well defined by a
power law,
$$ {\xi (r) = \bigg ({r_o \over r}\bigg)^\gamma,}
\eqno(20)$$
where $r_o=3.76$ h$^{-1}$ Mpc and $\gamma = 1.66$ for IRAS
(Fisher et. al I), while $r_o = 5.4$ h$^{-1}$ Mpc and $\gamma = 1.8$ for
the UGC (DP83). The integral is
easily evaluated numerically. The
N-body case is more complicated because the correlation
function is poorly fit by a power law and has to be treated
with more care. The correlation function $\xi(r)$ for the full 
N-body simulation was measured and 
used in the evaluating the integral Eq (19) numerically.

Before estimating the intrinsic dispersion $\sigma_I$, we found it
necessary to first subtract the
average of the measured $D(\Delta v)$ for the range $1000 < \Delta v < 1200$ 
km/s from the full distribution $D(\Delta v)$. 
This correction removes small biases that affect the measured
$G(\Delta v)$ for each galaxy (Eq. 16) by our requirement that each
galaxy have neighors in excess of random; fluctuations of randomly distributed
background galaxies can, on occasion, populate a long tail in the $G(\Delta v)$
distribution and would bias the inferred dispersion.
For a given  $\sigma_I$, we can construct model distribution function
$M(\Delta v)$, again
with $\bar M(1000 < \Delta v < 1200)$ subtracted from the distribution.
We shall denote subscript {\it c} for background subtracted distributions.
We furthermore adjust the model normalization $C_1$ such that
$$ \int_0^{1000} D_c(\Delta v) d\Delta v = 
\int_0^{1000} M_c(\Delta v) d\Delta v \quad . \eqno(21)$$

To determine a best fitting model from these corrected distributions, 
we simply evaluate the sum of the squared deviation,
$$ \chi^2 = \sum_i \left( {D_c(\Delta v_i) - M_c(\Delta v_i) \over N_i}
\right)^2  \quad , \eqno(22)$$
where $N_i$ is the uncertainty of point $i$. We have chosen $N_i$ constant
but find little change in the best $\sigma_I$ for different weightings.

The measured velocity distribution is plotted in Figure 3 for each catalog,
along with the best fitting exponential and gaussian models;
also shown are
fits with $\sigma_I=0$.  The solid curve in each case is the same
as in Figure 1, apart from the adjustment of the zero point. 
The effects of the
velocity broadening are clearly observable.  
Results for the inferred value of $\sigma_I$
are listed in Table 4 for gaussian and exponential models, with the model
leading to the best $\chi^2$ given in boldface. 
The UGC catalog is best fit with an exponential model (i.e. the best $\chi^2$ 
is substantially 
lower), while the IRAS and N-body catalogs have lower $\chi^2$ when
using the gaussian velocity distribution.  It is not too surprising that
the real catalogs would behave this way, since the IRAS catalog undersamples
the richer regions of the UGC survey, and thus would be expected  to have
less of a tail in the distribution.  It is unclear why the N-body catalogs,
sampling a similar volume to the IRAS survey, 
do not similarly favor the exponential models.


 Note   that
the inferred  $\sigma_I$ does indeed drop by a factor of $\approx 2$ between
the cooled and uncooled simulation, as expected. This is a
demonstration that our statistical procedure is sensible, and the test
is not trivial since the selection of neighbors is performed in {\it redshift
space}. The uncooled N-body model
is so hot, however, that our estimate of $\sigma_I$ is compromised
by the relatively small window in velocity space from which we draw the 
neighbor pairs.  The derived $\sigma_1$ for the cooled N-body model, 192 km/s,
can be compared to the internal dispersion of the N-body models listed in
Table 2, $\approx 450$ km/s for pairs within
$2 h^{-1}$ Mpc.  Dividing this value by the same factors for the artificial
cooling and bias (2*1.19) leads to an expected $\sigma_1 = 189$ km/s, 
exactly as measured in redshift space.

The statistical errors given in Table 4 are the one sigma error contours 
derived from the $\chi^2$ procedure.  Assuming the best fit to be acceptable
over the twenty bins of data with two degrees of freedom
($\sigma_I$ and the overall normalization), we list
the change in $\sigma_I$ that increments $\chi^2$ by 1/18 of its minimum  
value.

An independent estimate of the errors can be derived by
a Monte Carlo experiment.   We have constructed five independent mock IRAS
catalogs from the same large N-body simulation and have cooled their
velocity fields by the method described by Davis \etal 1996.  Each of
these simulations have similar size and sampling density to the IRAS
catalog. Comparison of  the individual estimates of $\sigma_I$ for these
mock catalogs simulates all the statistical effects of our procedure, 
including sampling variance. The dispersion of the best
$\sigma_I$ measured from the five mock catalogs is only
30 km/s, or 9\%, which is consistent with the errors listed in Table 4.

In Table 4, we also list
 $\sigma_1$,   the final value of the single particle
one dimensional dispersion, which is simply $\sigma_I$ corrected for the 
selection bias
factor of 1.19 discussed above 
and by a factor of $\sqrt{2}$ to change from the
rms of the difference between two galaxies to the motion of a single
galaxy. 
Each $\sigma_1$ listed is based on the better of the gaussian/exponential
fits.
The numbers in boldface highlight the preferred models, 
either  exponential or gaussian,  that has the smaller $\chi^2$. 
The lower dispersion for the
IRAS galaxies is hardly a surprise, given that this catalog,
relative to optically selected
samples such as the UGC, undercounts
ellipitical and lenticular galaxies  that are most abundant in the dense,
hot cluster regions.  But note that even the UGC catalog leads to a dispersion
considerably less than that obtained from the {\it artificially cooled} N-body
models.

Our object weighted velocity dispersion is similar to
that defined by Rivolo and Yahil (1981).  The final quantity,
$\sigma_1$, is itself very similar to the mean velocity
dispersion within groups of galaxies (e.g. Nolthenius and White 1987;
Ramella, Geller, and Huchra 1989; Nolthenius, Klypin, and Primack 1994).  
In fact, the method
should yield virtually the same result because only galaxies in
groups have neighbors above the background level.  Our analysis is
distinct from those recent analyses that focus on the kinematics of dwarfs
around isolated large galaxies (e.g. Zaritsky and White 1994, Zaritsky \etal
1996). The information we have obtained tells us nothing about
 the dispersion versus
environment, as measured by e.g. Chengalur \etal (1996), Kepner \etal (1996).
However,
the derived $\sigma_1$ estimate should be robust and is designed specifically
for use in the filtered cosmic energy equation.

\bigskip
\centerline {\bigbf 4. Application of the Filtered Energy Equation}

Given the estimate of $\sigma_1$, we have all the ingredients to apply the
filtered cosmic energy equation (15).  As discussed above, a suitably filtered
cosmic potential energy can be derived by simply limiting the integration range
of $J_2$.  Consider a sample $j$ and suppose that its
effective bias relative to the mass distribution is $b_j$ (i.e.  we define $b_j$
by the expression $J_{2,j} = b_j^2 J_{2,{\rm mass}}$).  There is a slight
ambiguity in determining the actual filtering scale associated with the cylinder
used to determine $\sigma_1$, and to avoid incurring bias associated with this
uncertainty, since we know that $\Omega _{N-body}=1$, $b_{N-body} =1$, we can
examine the ratio,
$$ \left({\Omega_j\over \Omega_{N-body}}\right)
\left(b_{N-body} \over b_j\right)^2 = \tilde\Omega = {\sigma_{1,j}^2\over
\sigma_{1,N-body}^2} {J_{2,N-body}\over J_{2,j}} \eqno(23)$$ 
where we define an
effective density parameter $\tilde\Omega \equiv {\Omega_j / b_j^2}$.  
The ratio
(23) allows us to eliminate from our estimate of $\tilde\Omega$ any dependence
on the relative filtering scales of $\sigma_1$ versus $J_2$, as well as the
appropriate value of $\tilde g$, but is biased
if the scale dependence in  $\tilde g$ is not the same in both the
N-body and real catalogs. The other uncertain parameters are $r_{max}$,
the upper limit, and $r_{min}$, the lower limit for evaluation of the $J_2$
integral.  Although $J_2$ converges at $r=0$, 
we choose $r_{min}=0.1$ h$^{-1}$ Mpc, roughly the separation of the
closest pairs considered, so as to eliminate from analysis the velocity
dispersion internal to galaxies and to consider only the dispersion of galaxies
moving relative to each other.  The appropriate $r_{max}$ must be in the range
$2-10h^{-1}$ Mpc to match the velocity filtering, but fortunately the exact
choice of $r_{max}$ is not critical.  In Table 5, this ratio is listed as a
function of $r_{max}$ for maximum projected pair separation of $2$ h$^{-1}$ Mpc,
using the values of $\sigma_1$ derived from Table 4.  We find that the measured
value of $\tilde\Omega$ is fortunately reasonably insensitive to variations in
$r_{max}$ for both the IRAS and UGC data.  With an N-body simulation that better
matched the $\xi(r)$ of observed catalogs, there would ideally be no sensitivity
to the uncertain value of $r_{max}$.

The statistical precision of our estimate of $\tilde\Omega$ is determined by the
statistical errors of $J_2$ and of $\sigma_1$.  Fisher etal (1994) estimate that
$\sigma_8$ for the IRAS 1.2Jy catalog has a statistical error of 6\%.  We shall
assume that $J_2$ has a similar error.  The dominant statistical error is
uncertainty in the estimate of $\sigma_1$, which leads to a statistical
precision of 34\% for the measurement of $\tilde\Omega$ for the IRAS and UGC
samples.  Larger, denser samples of galaxies should enable one to greatly
improve the quality of the $G(\Delta v)$ distribution and to greatly reduce the
statistical uncertainty of $\tilde\Omega$.


Table 5 shows that both the IRAS and UGC catalogs lead to estimates
$\tilde\Omega \approx 0.14$.  Given the known $J_2(r)$ functions for
the IRAS and UGC surveys, one would have naively expected to find
$\tilde\Omega_{iras}/ \tilde\Omega_{ugc} = ( b_{ugc} / b_{iras} )^2 \approx
1.7$, where the latter follows from the known $\xi(r)$ and $\sigma_8$ values for
the two catalogs (Fisher \etal 1994).  The contrary result demonstrates that
both catalogs cannot be linearly biased tracers of the mass distribution.  In
fact, it is likely that neither sample is a linearly biased mass tracer on the
small scale we are probing in this analysis.

\bigskip
\centerline {\bigbf 5. Discussion and Conclusions}
\medskip

We have developed a single galaxy weighted statistic, $\sigma_1$, which contains
much of the same information as the traditional pair weighted statistic,
$\sigma_{12}(r)$, but we believe the new statistic will prove to be much more
stable than $\sigma_{12}(r)$.  We have computed a galaxy weighted measure of the
galaxy pair dispersion for a semi-volume limited catalogs of 2374 IRAS galaxies
and 1959 UGC galaxies.  For the IRAS sample, we find a value for the
one-particle one-dimensional velocity dispersion of $\sigma_1 =96
\pm 16$ km/sec based on analysis of galaxy pairs with projected separation, $r_p
< 2$ h$^{-1}$ Mpc and line-of-sight separation less than or equal to 1000
km/sec, while our best estimate of $\sigma_1$ for the UGC sample is $130 \pm 15$
km/s.  The higher value for UGC is not unexpected, given that it better samples
the hot centers of clusters of galaxies than does the IRAS survey.  These
velocity dispersions are appropriate for use in a filtered version of the
Layzer-Irvine equation and lead to an estimate of the density parameter,
$\tilde\Omega \equiv \Omega/b^2$.  As usual there is a degeneracy between bias
in the galaxy tracer and $\Omega$.  Our best estimate for both the IRAS and UGC
catalog is $\tilde\Omega \approx 0.13- 0.17$, consistent with some recent
determinations on this scale (Fisher \etal 1994, Carlberg \etal 1996, Fisher and
Nusser 1996).  It is encouraging that the results for $\tilde\Omega$ obtained
using the more densely sampled optical catalog seem to be roughly consistent
with those obtained using the IRAS galaxies, in spite of the differences in
their $J_2$ integrals, but at the same time, this consistency shows the
inadequacy of the naive concept of linearly biased tracers of the mass field.

Our results clearly demonstrate that $\Omega=1$ N-body models, normalized with
mass correlation functions close to that observed in the galaxy distribution,
have a one particle rms velocity that is more than twice that observed in the
galaxy distribution.  The particle mass of the nbody models used here is
$6~10^{12} \Msun$, so that internal velocities within individual galaxies do not
significantly affect our estimate of $\sigma_1$.  Note also that $\sigma_1$ for
the cooled N-body catalogs, when multiplied by 2 to correct for the cooling and
by 1.19 to reinsert the bias factor, is very close to the internal velocity
dispersion of 450 km/s listed in Table 2 for the small scale velocity dispersion
of the real space (uncooled) N-body catalog.  It thus appears that our estimator
of $\sigma_1$ is capable of an accurate measurement of $\sigma_1$ in the range
$0-300$ km/s.  For optimal determination of larger values of $\sigma_1$, a
larger limit to $\Delta v$ for the pair counts would be appropriate, but this is
only possible for larger, deeper redshift catalogs.

This statistic has a significant advantage over $\sigma_{12}(r)$ in
that it weights each galaxy equally and is not dominated by the abundant pairs
in rich cluster centers, where the velocity dispersion is
known to be much higher than average. The statistic should be much more
robust to the inclusion of rare clusters than is $\sigma_{12}(r)$.  Given
that the $J_2$ integral on small scales is known to be very stable from
one catalog to another, further estimates of $\sigma_1$, combined with
 the filtered cosmic energy equation, should lead to a reliable
 estimate of the effective $\tilde\Omega$ on the scale of a few Mpc.

The method we describe for estimation of $\sigma_1$ has the disadvantage
that we have deleted the majority of galaxies from consideration because
they have too few neighors.  
Variations on our procedure might well turn out
to be more suitable for larger catalogs.  
However, preliminary estimates (Davis \etal 1997) derived
from the large LCRS survey (Shectman \etal 1996) indicate that this
problem is overcome with the dense sampling of LCRS; most galaxies
have sufficient neighbors in this densely sampled survey to be included
and the derived $\sigma_1$ is very stable.

The ``coldness" of the typical thermal environment of galaxies has
been noted for quite some time (Peebles 1992; Ostriker and Suto
1990). Most previous estimates of the small
scale velocity field have been based on the observed  pair
dispersion $\sigma_{12}$, which is perhaps suspect.  
  The results presented here imply that the low rms velocity of galaxies
relative to their neighbors is 
not simply a problem of the inclusion  or absence of a sufficient
number of rare, rich clusters. The average galaxy in a typical group
has a local velocity field that is much colder than expected in high 
$\Omega$ models; the mock
catalogs demonstrate the problem very clearly. 
Further evidence for a cold velocity field comes from Tully-Fisher distance
estimates using such  sophisticated statistical
procedures as the VELMOD algorithm described by Strauss and Willick
(1995).   Willick \etal (1996) suggest a very 
similar small scale random velocity, 
$\sigma_1 \approx 125\pm20$ km/s,
although the weighting and filtering of their estimate are rather different
from those of the measure we present here.

Another argument for cold velocity fields is the remarkably quiet Hubble flow in
the vicinity of the Local Group of galaxies; the rms peculiar velocity of
galaxies within $5h^{-1}$ Mpc of the Milky Way is only 60 km/s (Schlegel \etal
1994) This result is extremely difficult (Schlegel \etal 1994), if not
impossible (Govertano \etal 1996) to reconcile with any N-body simulation for an
open or closed Universe!  Understanding why the observed small scale velocity
field is so cold remains one of the major unsolved mysteries of large scale
structure.

The quantity measured in our analysis, ${\Omega / b_j^2}$, is formally very
nearly equal to the square of $\beta \equiv {\Omega^{0.6} / b_j}$, the
quantity measured in the analyses of large scale flows; however, this
analogy presumes that the bias factor $b$
has the same meaning on these very different scales.  The IRAS results quoted
here are consistent with $\beta \approx 0.30-0.35$, assuming a scale-independent
bias.  This value is somewhat lower than that derived from recent comparisons of
observed peculiar velocities and the IRAS predicted gravity field (Davis, Nusser
and Willick 1996; Willick \etal 1996).  This modest difference between a
measurement on scales of 1-2$h^{-1}$ Mpc, and another with an effective
scale in the range 10-50$h^{-1}$ Mpc, is perhaps the signature of
scale-dependent bias, as discussed recently by Kauffmann \etal (1996).
A modest trend of increasing $\Omega$ estimates with
measurement scale would also be expected in mixed dark matter models with
$\Omega_\nu \approx 0.2-0.3$ (Primack 1994; 1996) in which there is some
suppression of growth within galaxy-sized halos; on the scale of the 
large-scale flows, the dark matter should fully
participate in the clustering and the measured $\beta$ should reach its
asymptotic limit.  

The cold thermal environment of the local galaxy distribution 
implies either that the galaxies are a strongly
biased tracer on small scales and that the mass normalization is 
$\sigma_8 < 0.4$ in an
$\Omega=1$ model, or that $\Omega \approx 0.15-0.4$ with the galaxies
roughly tracing the mass distribution, perhaps with
modest bias on small scale.  The former conclusion is rather
far from the COBE suggested normalization in many (although not all) 
models of large scale
structure (Bunn, Scott, and White, 1995), and the latter suggestion,
while consistent with many observations, is not consistent with the
higher values derived from the POTENT analysis on larger scale (Dekel
etal 1993), nor is it consistent with  the naive expectations from
inflation.  Neither $\Omega=1$ and $\sigma_8\leq 0.4$ nor $\Omega\leq
0.25$ and $\sigma_8=1$ appears consistent with the observed abundance
of rich galaxy clusters (White et al 1993). 
Low values of $\Omega$ have the  additional problem of an
overly steep mass auto-correlation function, which requires 
galaxies to be  anti-biased ($b < 1$) on small scales (e.g. Davis et
al 1985; Klypin, Primack,
\& Holtzman 1995; Cole \etal 1997), 
a difficulty which has been known but neglected for
years.  Such an antibias will be difficult to reconcile with the
already low values of $\tilde\Omega$ derived here.

\bigskip
\centerline{\bf ACKNOWLEDGEMENTS}

We acknowledge helpful conversations with M. White, S. Zaroubi, and M. Craig.
This work was supported in part by NSF grant AST95-28340.

\np

\medskip
\noindent {\bf Table 1}~~ $K/|W|$ from nbody simulations
$$\vbox{\offinterlineskip\halign{
\strut#&\vrule#\quad&
\hfil$#$\hfil&
\quad\vrule#\quad&
\hfil#\hfil&\quad\vrule#\quad&
\hfil#\hfil&\quad\vrule#\quad&
\hfil#\hfil&\quad\vrule#\quad&
\hfil#\hfil&\quad\vrule#\cr
\noalign{\hrule}
&&&&&&$\Omega=1$, n=--1&& $\Omega=0.1$, n=--1&&$\Omega=1$, CDM& \cr
\noalign{\hrule}
&&$${\sigma_s / r_0}$$&&${K_{small}}$&&
${(K/W)_{small/large}}$&& ${(K/W)_{small/large}}$&& ${(K/W)_{small/large}}$& \cr
\noalign{\hrule}
&&.6&& 7.8&&0.78/0.58&&0.71/0.36&&0.92/0.42& \cr
&&1.2&& 8.9&&0.73/0.63&&0.65/0.41&&0.79/0.48& \cr
&&2.4&& 10.6&&0.73/0.66&&0.63/0.42&&0.75/0.52& \cr
&&4.8&&12.5&&0.73/0.67&&0.62/0.42&& & \cr
\noalign{\hrule}
}}$$

\medskip
\noindent {\bf Table 2}~~ Scale dependence of internal and external dispersion 
of mock catalog
$$\vbox{\offinterlineskip\halign{
\strut#&\vrule#\quad&
\hfil$#$\hfil&
\quad\vrule#\quad&
\hfil#\hfil&\quad\vrule#\quad&
\hfil#\hfil&\quad\vrule#\quad&
\hfil#\hfil&\quad\vrule#\quad&
\hfil#\hfil&\quad\vrule#\quad&
\hfil#\hfil&\quad\vrule#\cr 
\noalign{\hrule}
&&\omit Scale (Mpc)&& 
       ${\sigma (\Delta v_p)}$&&$ {\sigma (\bar v_p)}$&&\bf $\sigma_{rms}$
       &&fraction&&
       $\sigma_{\rm with\ neighbors}  \over \sigma_{\rm total}$\cr
&&{\rm radius}&&\omit{\rm internal}&&\omit{\rm 
external}&&{}&&{}&&{}&\cr 
\noalign{\hrule}
&&1&& 458&& 420&& 667&& 0.29&& 1.23& \cr
&&2&& 451&& 407&& 640&& 0.43&& 1.18& \cr
&&3&& 436&& 389&& 604&& 0.56&& 1.11& \cr
&&4&& 425&& 372&& 577&& 0.69&& 1.06& \cr
&&5&& 418&& 365&& 562&& 0.80&& 1.04& \cr
&&{\rm all\ particles}&&{}&&{}&&541& \cr
\noalign{\hrule}
}}$$


\bigskip
\noindent {\bf Table 3}~~ Fractions with excess neighbors in the catalogs 
(redshift space)
$$\vbox{\offinterlineskip\halign{
\strut#&\vrule#\quad&
\hfil$#$\hfil&
\quad\vrule#\quad&
\hfil#\hfil&\quad\vrule#\quad&
\hfil#\hfil&\quad\vrule#\cr
\noalign{\hrule}
&&\omit Survey&&
       {fraction with neighbors}
&&$\sigma_{\rm with\ neighbors}  \over \sigma_{\rm total}$&\cr
\noalign{\hrule}
&&\omit N-body&&0.28&&1.19&\cr
&&\omit IRAS&&0.25&&{}&\cr
&&\omit UGC&&0.29&&{}&\cr
\noalign{\hrule}
}}$$
\medskip

\medskip
{\baselineskip=20pt
\noindent {\bf Table 4 }~~ Dispersion Results (exponential \& gaussian  models)
$$\vbox{\offinterlineskip\halign{
\strut#&\vrule#\quad&
\hfil$#$\hfil&
\quad\vrule#\quad&
\hfil#\hfil&\quad\vrule#\quad&
\hfil#\hfil&\quad\vrule#\quad&
\hfil#\hfil&\quad\vrule#\quad&
\hfil#\hfil&\quad\vrule#\cr
\noalign{\hrule}
&&&&$\sigma_I$ (exponential) &&$\sigma_I$ (gaussian) && $\sigma_1$&\cr
\noalign{\hrule}
&&{\rm IRAS}&&180${+35\atop -20}$ &&${\bf 160{+15\atop -30}}$&&{ 95$\pm$16} &\cr
&&{\rm UGC}&& ${\bf 220{+30\atop -20}}$&& $180 {+25\atop -20}$ &&{ 130$\pm$15} &\cr
&&{\rm N-body} && $1000 {+200\atop -90}$&&${\bf 540 {+80\atop -45}}$&&${ 325 
{+40\atop -27}}$ &\cr
&&{\rm N-body cooled}&& $500 {+25\atop -45}$&& {\bf 320 $\pm$ 10}&& 
{ 190 $\pm$ 10} &\cr
\noalign{\hrule}}}$$  }
\medskip

\noindent {\bf Table 5}~~ $\tilde\Omega$ versus $r_{max}$
$$\vbox{\offinterlineskip\halign{
\strut#&\vrule#\quad&
\hfil$#$\hfil&
\quad\vrule#\quad&
\hfil#\hfil&\quad\vrule#\quad&
\hfil#\hfil&\quad\vrule#\quad&
\hfil#\hfil&\quad\vrule#\cr
\noalign{\hrule}
&&		r_{max}~(h^{-1} Mpc)  &&  IRAS  &&  UGC& \cr
\noalign{\hrule}
&&		2     && 0.17	&&  0.14 & \cr
&&		4     && 0.14	&&  0.13 & \cr
&&		6     && 0.13	&&  0.12 & \cr
\noalign{\hrule}}}$$


\np
\centerline {\bf References}
\medskip

\noindent Bean A. J., Efstathiou G., Ellis R. S., Peterson B. A.,
Shanks T. 1983, MNRAS, 205, 605

\noindent Bartlett, J.G., \& Blanchard, A. 1993, in ``Cosmic Velocity
Fields: IAP 1993", ed.  F. Bouchet \& M Lachi\`eze-Rey, Editions Frontieres,
p. 281.

\noindent Bartlett, J.G., \& Blanchard, A. 1996, A\&A, 307, 1

\noindent Brainerd, T.G., Bromley, B.C., Warren, M.S., \& Zurek, W.H. 1996,
Ap.J. 464, L103

\noindent Bunn, E. F., Scott, D., \& White, M. 1995, Ap.J. 441, L9

352, L29.

\noindent Carlberg, R, Yee, H., Ellingson, E., Abraham, R., Gravel, P., 
Morris, S., \& Pritchet, C. J. 1996, Ap.J. 462, 32

\noindent Chengalur, J.N., Salpeter, E.E., \& Terzian, Y. 1996, ApJ., 461, 546

\noindent Cole, S., Weinberg, D.H., Frenk, C.S., \& Ratra, B. 1997, 
 astro-ph/9702082
 
\noindent Davis M., Geller M., and Huchra J. P. 1978,  ApJ, 221, 1

\noindent Davis M. and Peebles P. J. E. 1977, ApJS, 34, 425

\noindent Davis M. and Peebles P. J. E. 1983, ApJ, 267, 465. (DP83)

\noindent Davis M., Efstathiou G., Frenk C. S. and White S. D. M. 1985, 
Ap J 292, 371-394.
 
\noindent Davis M., Nusser A., and Willick 1996, Ap.J., 473, 22

\noindent Davis, M., Lin, H., \& Kirshner, R. 1997, in preparation

\noindent Dekel A., Bertschinger E., Yahil A., Strauss M.A., Davis M., 
Huchra J. P. 1993, Ap J, 412, 1-21.

\noindent de Lapparent V., Geller M., and Huchra J. 1988, ApJ, 332,
44

\noindent Fall S. M. 1976, MNRAS, 172, 23p

\noindent Fisher K. B. 1992, PhD thesis, University of California, Berkeley

\noindent Fisher K. B., Davis M., Strauss M. A., Yahil A., Huchra J.
P. 1994, MNRAS, 266, 50 (Fisher et. al. I)

\noindent Fisher K. B., Davis M., Strauss M. A., Yahil A., and
 Huchra J. P. 1994,
MNRAS, 267, 927 (Fisher et. al. II)

\noindent Fisher K. B., Huchra J. P., Strauss M. A., Davis M., Yahil A.,
 and Schlegel D. 1995, Ap J Supp., 100, 69-103.

\noindent Fisher K., and Nusser A. 1995, preprint astro-ph/9510049

\noindent Gelb  J. M. and Bertschinger E. 1994, Ap J. 436, 467-490.

\noindent Gelb  J. M. and Bertschinger E. 1994, Ap J. 436, 491-508.

\noindent Geller M., and Peebles P. J. E. 1973, Ap J., 184, 329

\noindent Govertano, F., Moore, B., Cen, R., Stadel, J., Lake, G., and Quinn, 
T., 1996, astro-ph/9612007

\noindent Guzzo L., Fisher K., Strauss M., Giovanelli R. Haynes M. 1996,
Ap.J., 463, 395

\noindent Hale-Sutton D., Fong P., Metcalfe N., Shanks T. 1989, MNRAS, 237,
	569

\noindent Juszkiewicz R, and Yahil A. 1989, Ap J. Lett., 346, L49.

\noindent Kauffmann G., Nusser A., and Steinmetz M. 1995, preprint 
astro-ph/9512009.


\noindent Kepner, J., Summers, F., and Strauss, M. 1996, preprint 
astro-ph/9607097 

\noindent Klypin, A., Primack, J., \& Holtzman, J. 1995, preprint 
astro-ph/9510024

\noindent  Marzke R.O., Geller M.J., da Costa L.N., Huchra J.P. 1995, 
Astron. J, 110 (no.2), 447-501.

\noindent Mo H. J., Jing Y. P., and Borner G. 1993, MNRAS, 284, 703.

\noindent Mo, H. J., Jing, Y. P., and Borner, G. 1996, 
preprint astro-ph/9607143

\noindent Nolthenius, R, and White, S. D. M. 1987, MNRAS, 22, 505

\noindent Nolthenius, R., Klypin, A., and Primack, J. R. 1994, Ap.J. Lett.,
	422, L45

\noindent Ostriker J. P. and Suto Y. 1990, Ap J, 348, 378-82.



\noindent Peebles P. J. E. 1979, Astron. J, 84, 6, 730

\noindent Peebles P. J. E. 1980 ``The Large-Scale Structure of the Universe".
Princeton University Press, Princeton, NJ

\noindent Peebles P. J. E.  1981 Ap J, 248, 885-97.

\noindent Peebles P. J. E., 1992, in ``Relativistic Astrophysics and Particle
Cosmology", Texas/PASCOS 92 Symp. ed. C. W. Akerlof \& M. A. Srednicki
(Ann. NY Acad. Sci., vol 688), 84


\noindent Primack, J. R. 1994, in ``Proceedings of the International
School on Cosmological Dark Matter", ed. J. Valle, A. Perez, World Scientific,
p. 81.

\noindent Primack, J.R. 1996, astro-ph/9610078

\noindent Ramella, M., Geller, M. and Huchra, J. 1989, ApJ., 344, 57

\noindent Rivolo A. R., and Yahil A. 1981, ApJ, 251, 477

\noindent Santiago B., Strauss M., Lahav O., Davis M., Dressler A.,
\& Huchra J. 1995, Ap.J., 446, 457

\noindent Santiago B., Strauss, M. A., Lahav, O., Davis, M. Dressler, A.,
and Huchra, J. 1996, Ap.J., 461, 38

\noindent Schlegel, D, Davis, M., and Summers, F. 1994, Ap.J. 427, 527

\noindent Shectman, S., Landy, S.D., Oemler, A., Tucker, D.L., Lin, H., 
Kirshner, R.P., \& Schechter, P.L. 1996, astro-ph/9604167

\noindent Somerville, R., Davis, M., \& Primack, J. 1996, astro-ph/9604041

\noindent Strauss M., Yahil A., Davis M., Huchra J. 1992, ApJ 397,
395

\noindent Strauss, M., \& Willick, J. 1995, Phys. Reports, 261, 271.

\noindent White, S.D.M., Efstathiou, G. and Frenk, C.S. 1993 MNRAS
262, 1023

\noindent Willick, J., Strauss, M., Dekel, A., and Kolatt, T. 1996,
astro-ph/9612240

\noindent Zaritsky, D. and White, S.D.M. 1994, Ap.J. 435, 599

\noindent Zaritsky, D., Smith, R., Frenk, C., \& White, S 1996, 
astro-ph/9611199

\noindent Zurek W. H., Quinn P. J., Salmon J. K., 
Warren M. S., 1994, ApJ, 431, 559






\np
\myfig {1} {20.0}
{\includegraphics{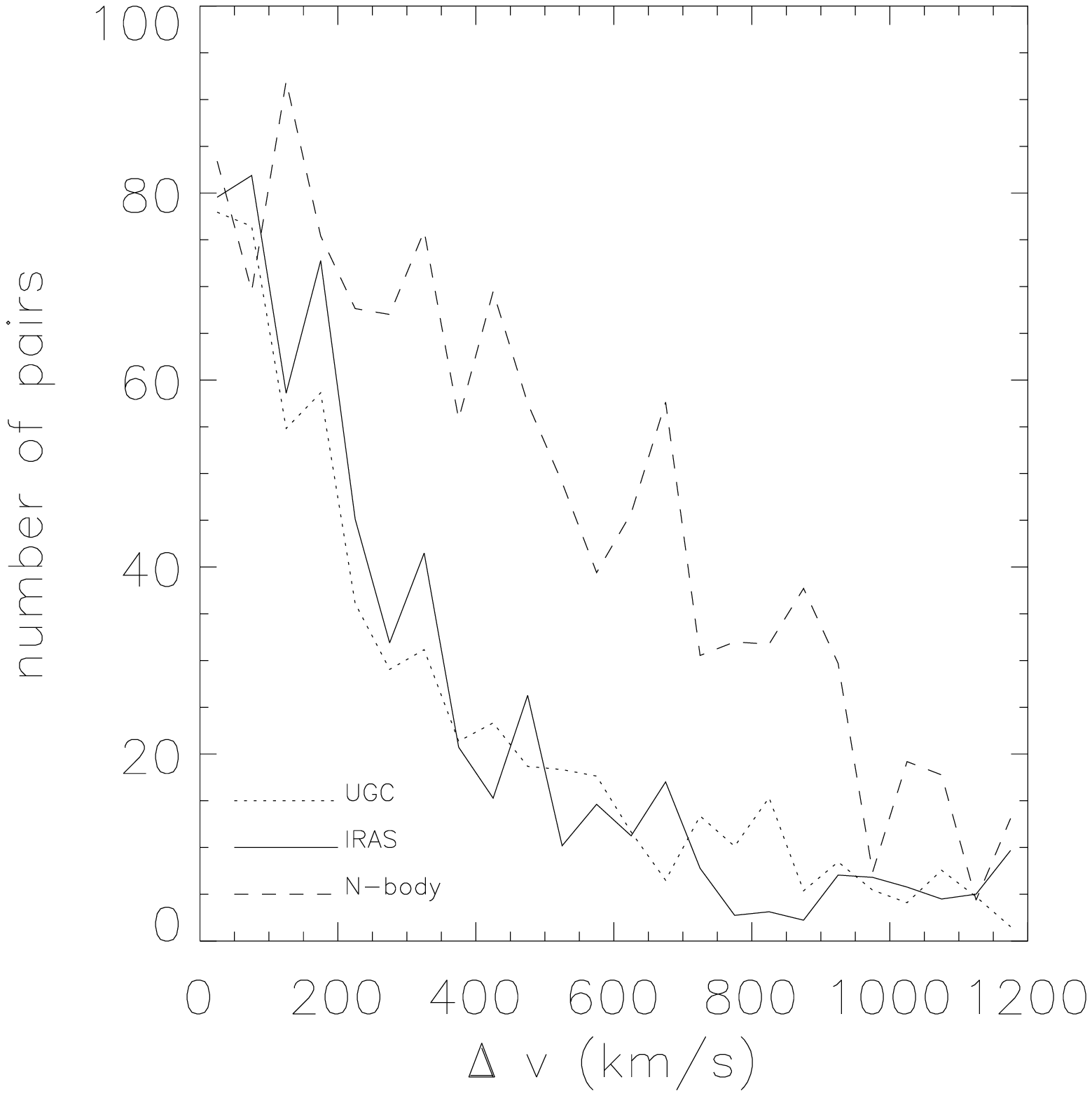}}
{\noindent The distributions of the number of pairs within
a cylindrical radius of $2h^{-1}$ Mpc as 
a function of velocity separation of the pair (galaxy-weighted). }

\np
\myfig {2} {20.0}
{\includegraphics{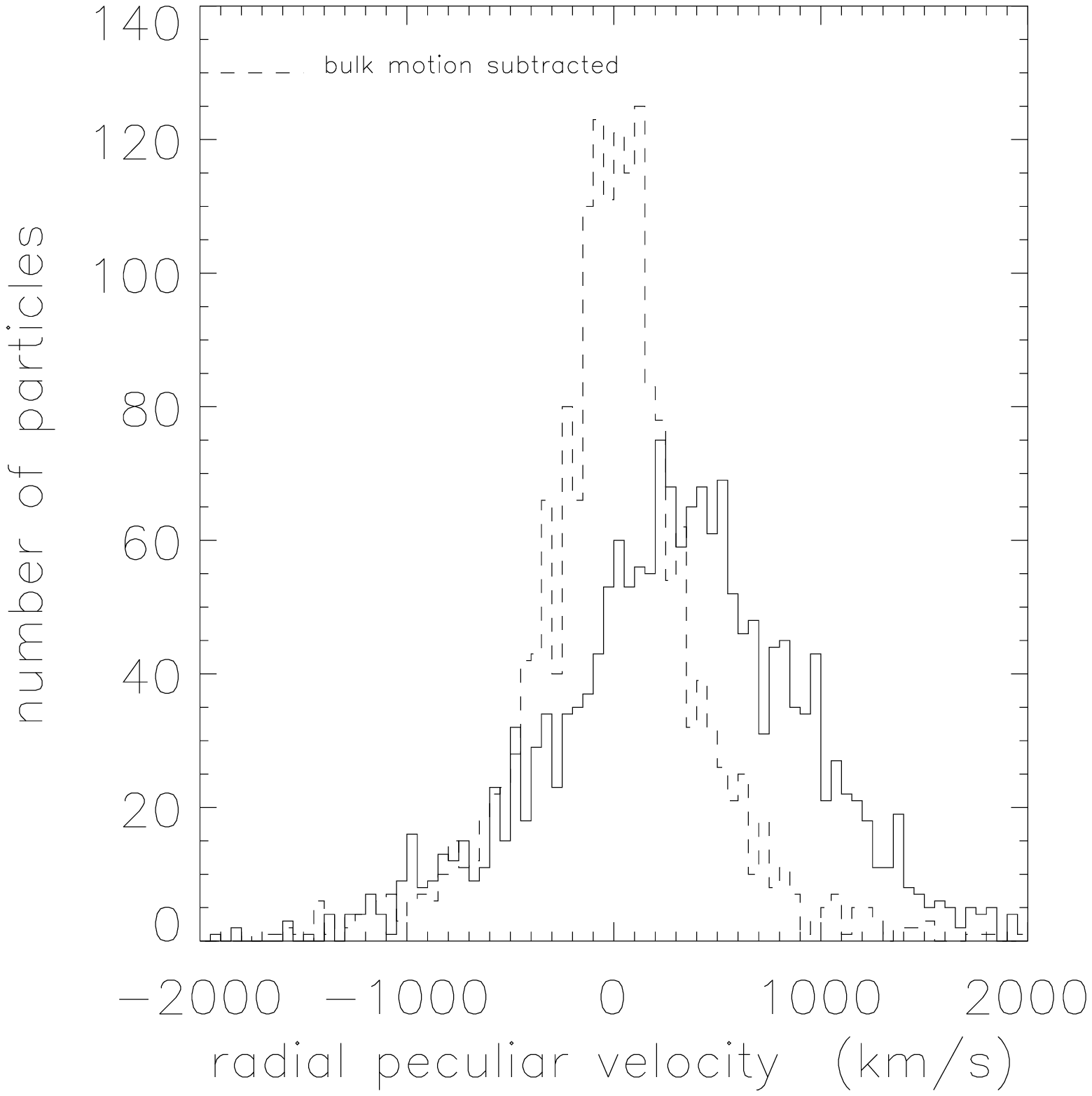}}
{The radial velocity distribution function
of N-body points having  pairs with  
real space separation $r_p<2$ h$^{-1}$ Mpc.  The solid curve is the
peculiar velocity relative to the center of the mock catalog. The dashed
curve is the velocity of the point relative to the mean velocity of
all points within a neighbor radius $r_p = 2h^{-1}$ Mpc.
}

\np
\myfig {3} {20.0}
{\includegraphics{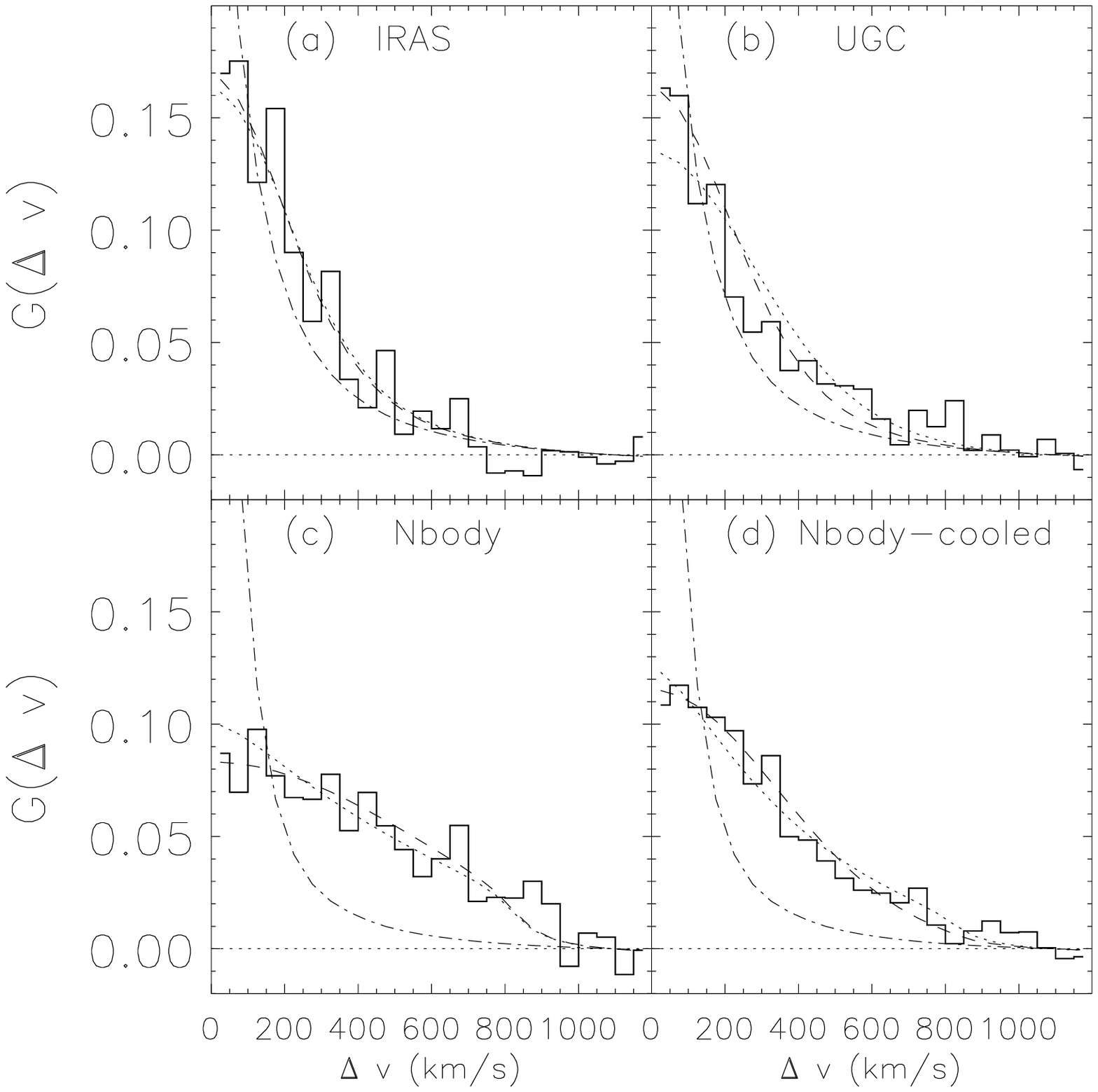}}
{  The observed distributions $G(\Delta v)$
 plotted as solid curves
 along with both the best fitting Gaussian (dotted curve) and  
 exponential models (dashed curve).  The dot-dashed curve is a model with
 $\sigma_I = 0$.
 (a) IRAS (b) UGC (c) N-body (d) N-body cooled}

\end